\documentclass[twocolumn]{aastex701}
\usepackage{amsmath}
\usepackage{afterpage}

\newcommand{\vmax}{v_{\rm max}}
\newcommand{\amax}{a_{\rm max}}
\newcommand{\thA}{\theta_{\rm A}}
\newcommand{\DA}{D_{\rm A}}
\newcommand{\Ds}{D_{\rm s}}
\newcommand{\dF}{d_{\rm F}}
\newcommand{\db}{d_{\rm b}}
\newcommand{\ms}{m_{\rm s}}
\newcommand{\PA}{P_{\rm A}}
\newcommand{\Liso}{L_{\rm iso}}
\newcommand{\taueff}{\tau_{\rm eff}}
\newcommand{\fb}{f_{\rm b}}
\newcommand{\GamL}{\Gamma_{6{\rm h}}}
\newcommand{\kms}{\,{\rm km\,s^{-1}}}
\newcommand{\um}{\,\mu{\rm m}}

\defcitealias{gl15}{GL15}
\shorttitle{Optical light-sail leakage with \textit{Roman}, \textit{Rubin} and \textit{Euclid}}
\shortauthors{Guillochon \& Loeb}

\begin{document}

\title{Optical Flashes from Beam-Driven Light Sails with the \textit{Roman}, \textit{Rubin} and \textit{Euclid} Observatories}

\correspondingauthor{James Guillochon}

\author[0000-0002-9809-8215]{James Guillochon}
\affiliation{Indepedent Researcher}
\email[show]{guillochon@gmail.com}

\author[0000-0003-4330-287X]{Abraham Loeb}
\affiliation{Astronomy Department, Harvard University, 60 Garden Street, Cambridge, MA 02138, USA}
\email{aloeb@cfa.harvard.edu}

\begin{abstract}
The primary challenge of rocket propulsion is the burden of accelerating the
spacecraft's own fuel. Light sails leave the propellant at home, with the achievable
speed set by the sail area, the thermal tolerance of its material, and the power of the
driving array. In \citet{gl15} we showed that leakage from a microwave array driving
such a sail between habitable worlds produces Jansky-level radio transients lasting
tens of seconds at $100$~pc. We take that leak to optical and near-infrared,
where Fresnel matching would
shrink the aperture to $\sim\!20$--$100$~m at $1\,\mu{\rm m}$, but the intensity on that
aperture is $2\times10^{8}\,{\rm W\,m^{-2}}$, five orders above published
directed-energy loadings. Spreading $1.5$~TW at $\sim20\,{\rm kW\,m^{-2}}$ is a
$\sim10$~km phased array, pushing the emitters; Lubin's
$10^{3}\,{\rm W\,m^{-2}}$ loading is a $40$~km array, comfortable but more expensive.
A tenth-wave delay on $40$~m tiles sized so one tile still
covers the sail leaks $0.054$ of the
power ($80$~GW) into a halo. The typical Galactic
detection is a single $\sim30$~s peak at $m_{\rm AB}\simeq19.4$ in F146
at $8$~kpc; a pair $87$~s apart, about one in five, is the confirmation test.
\emph{Roman}'s Galactic Bulge Time Domain Survey reaches
$109$~kpc at $8\sigma$ on a faint host, so a pointed flash of that halo is visible
from anywhere in the Galaxy if the beam points at us. For $10$--$40$~km optical beamers
held to $\sim\lambda/10$ at $v_{\rm max}\gtrsim250\,{\rm km\,s^{-1}}$, $N_{\rm det}=1$ wants
$\Gamma_{6{\rm h}}\simeq16$ and a null search limits $\Gamma_{6{\rm h}}\lesssim49$, a factor-of-three
window in launch rate. The real
threshold is set by PSF-coincident stellar and instrumental transients. If intensity-limited optical beamers are
commonly employed in our galaxy, this activity could be revealed by \emph{Roman},
\emph{Rubin} and \emph{Euclid} at no additional observing cost.
\end{abstract}


\section{Introduction} \label{sec:intro}

Travel time between habitable worlds by chemical rocket is long, $\sim8$--$9$
months one way for an Earth--Mars Hohmann transfer and $\sim2$~yr for the
round trip once the wait for the return window is included, and the rocket
equation exacts an exponential penalty in propellant mass for every increment
in speed. Beam-driven light sails leave the propellant at home, with the
achievable speed set by the sail area, the thermal tolerance of its material, and the
power of the driving array \citep{forward84,benford13,lubin16,parkin18}. The required
beam powers are large, terawatts for tonne-class payloads at $1$~gee, so the leakage
of radiation past the sail is a technosignature with a well-posed physical model. In
\citet{gl15}, hereafter \citetalias{gl15}, we constructed that model for a microwave
array conducting Earth--Mars transits, and found that the optimal frequency for a
$1.5$~km aperture lies near $68$~GHz, producing transients of a few Jansky at
$100$~pc lasting tens of seconds. Beams that would drive a relativistic
interstellar sail have been identified with fast radio bursts \citep{lingam17}:
the observer sits in the main beam of a planet-scale emitter, and the optimal
frequency is a few GHz. That geometry is the beam itself, not the leak, and it
remains a radio search.

The same matching that sets the beam frequency also sets a Fresnel-matched
array diameter, and at optical wavelengths that aperture collapses from kilometers
to tens of meters. The cost analysis of \citet{benford13}, which \citetalias{gl15}
adopted and which favored microwaves, reflected high-power microwave sources in the
early 2010s. Directed-energy concepts for gram-scale to tonne-scale sails have since
been developed at optical wavelengths \citep{lubin16,atwater18,parkin18,lin25,gao24}, and
that is the branch being prototyped on Earth \citep{michaeli25,norder25}. A
civilization may well land in the optical or near-infrared because source
economics no longer favor microwaves, in which case radio searches are looking
in the wrong band. The same
shrinkage raises the intensity at the
aperture by the inverse area, and we find in \S\ref{sec:beamform} that power
handling, not Fresnel matching, is the binding constraint on the optical branch:
two envelopes, $10$~km and $40$~km, bracket published emitter loadings.

Three survey facilities now cover that window. The \emph{Nancy Grace Roman Space
Telescope} \citep{spergel15,akeson19,rotac25} was launched on 2026 August 30 and is
in cruise to L2 plus commissioning, with first images expected in early 2027. The
\emph{Vera C.\ Rubin Observatory} \citep{ivezic19} began the ten-year \emph{Legacy
Survey of Space and Time} (\emph{LSST}) on 2026 June 30. \emph{Euclid} is midway
through a $1.4\times10^{4}$~deg$^{2}$ survey \citep{laureijs11,scaramella22}; Quick
Releases Q1 (2025 March) and Q2 (2026 June 24) are public, with DR1-Foundation
scheduled for 2026 November. Between them these facilities cover the wavelength band of $0.3$--$2.3\um$
and will accumulate of order $10^{15}$ star-seconds of monitoring. They have been
discussed as technosignature platforms in general terms
\citep{davenport19,hambleton23,haqqmisra22}, but the specific transient of
\citetalias{gl15} has not been mapped onto them: the signal they can catch is not
a directed optical-maser beam \citep{schwartz61} or a nanosecond pulse \citep{howard04}
and not a narrow spectral line \citep{tellis17,zuckerman23}, but a tens-of-seconds, spatially unresolved,
single-band flash at the position of a star. The all-sky optical SETI experiment
\emph{PANOSETI} searches $0.35$--$1.65\um$ at nanosecond to second cadence with a
plate scale of $0.36$~deg per pixel \citep{wright18}; that experiment is built
for a collimated pulse, not for a tens-of-seconds PSF-shaped brightening of a star.

In \S\ref{sec:design} we generalize the design relations of \citetalias{gl15} to
arbitrary wavelength, show that a Fresnel-matched optical aperture cannot radiate
$1.5$~TW, and take two focused arrays as the leak we search for: $10$~km at
$\sim20\,{\rm kW\,m^{-2}}$ and $40$~km at the published $10^{3}\,{\rm W\,m^{-2}}$
loading.
\S\ref{sec:signal} describes the observable event. \S\ref{sec:surveys} assesses detection prospects with the
\emph{Roman}, \emph{Rubin} and \emph{Euclid} observatories, \S\ref{sec:backgrounds} treats
backgrounds, and \S\ref{sec:disc} concludes.

\section{Optical and near-infrared beamers} \label{sec:design}

\subsection{Design relations}

\begin{figure*}
\centering
\includegraphics[width=0.986\linewidth]{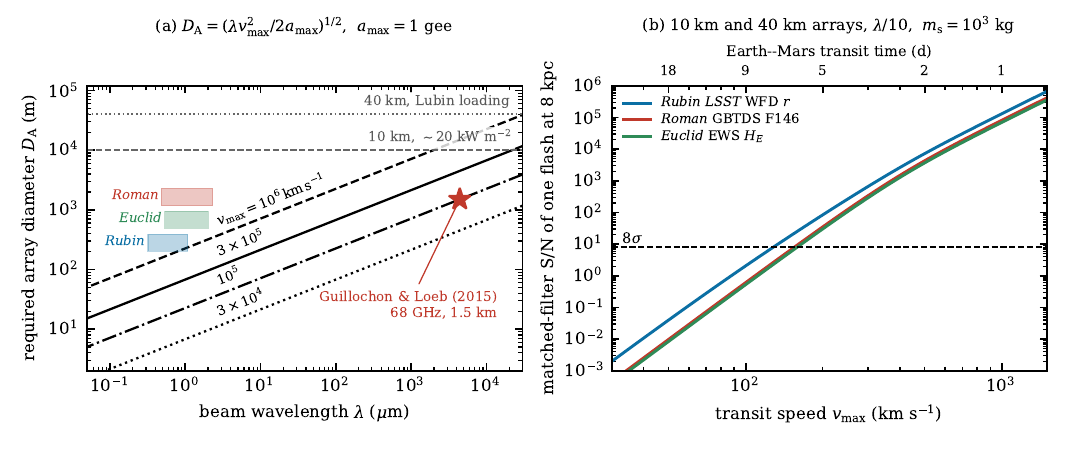}
\caption{(a) Array diameter required by the Fresnel-matching condition,
Equation~(\ref{eq:DA}), as a function of beam wavelength for four transit speeds at
$\amax=1$~gee. The star marks the $68$~GHz, $1.5$~km microwave design of
\citetalias{gl15}; the same mission conducted at $1\um$ requires a $23$~m aperture.
Labels along the loci are $\vmax$ in km~s$^{-1}$.
Horizontal lines mark the $10$~km ($\sim20\,{\rm kW\,m^{-2}}$) and $40$~km
(published-loading) envelopes of \S\ref{sec:beamform}.
Shaded boxes show the \emph{Rubin}, \emph{Euclid} and \emph{Roman} bandpasses.
(b)
Matched-filter signal-to-noise ratio of a single end-of-burn halo flash from a
$10$~km or $40$~km array of intercept-matched tiles with a $\lambda/10$ lock, for a system at
$8$~kpc and a $10^{3}$~kg payload at $1$~gee. The
upper axis converts $\vmax$ to an Earth--Mars transit time. The dashed line at
${\rm S/N}=8$ is the threshold adopted throughout: it is where the Gaussian
false-alarm expectation over the $1.7\times10^{14}$ read-trials of the GBTDS
falls to $\sim\!0.1$ events (\S\ref{sec:stat}). It is a floor set by noise
statistics. The real threshold will be set by the astrophysical and
instrumental backgrounds of \S\ref{sec:backgrounds};
Figure~\ref{fig:limits}c gives the yield surviving any harder threshold. All
three facilities cross it near $250\kms$ for a tonne-class sail at $1$~gee.}
\label{fig:design}
\end{figure*}

We adopt the design relations of \citetalias{gl15} and allow the wavelength
$\lambda$ to be a free parameter. A sail accelerated by an array of diameter
$\DA$ remains in the array's Fresnel zone, where the beam energy is delivered
through a constant cross-section rather than a constant solid angle, out to the
Fresnel length $\dF=\DA^{2}/\lambda$, and efficiency demands that an unfocused
burn end there, since beyond $\dF$ the wasted fraction grows rapidly. The terminal
speed is then $\vmax=(2\amax \dF)^{1/2}$, which fixes
\begin{equation}
\dF=\frac{\vmax^{2}}{2\amax}, \qquad
\DA=\left(\frac{\lambda \vmax^{2}}{2\amax}\right)^{1/2},
\label{eq:DA}
\end{equation}
which is Equation~(1) of \citetalias{gl15} solved for $\DA$ at fixed $\lambda$ rather
than for $\lambda$ at fixed $\DA$. The kinematic burnout
$\db=\vmax^{2}/(2\amax)$ coincides with $\dF$ only for that Fresnel-matched
aperture; a focused array can end the burn at $\db$ deep in the near field.
Writing
$\lambda_{1.06}\equiv\lambda/1.06\um$, $v_{400}\equiv\vmax/400\kms$,
$a_{1}\equiv\amax/1\,{\rm gee}$ and $m_{3}\equiv\ms/10^{3}\,{\rm kg}$, where
$\ms$ is the sail mass, and fixing $\vmax$, $\amax$ and $\lambda$ to those
fiducials we find
\begin{equation}
\DA = 93\,{\rm m}\,\lambda_{1.06}^{1/2}\,v_{400}\,a_{1}^{-1/2}.
\label{eq:DAnum}
\end{equation}
Figure~\ref{fig:design}a shows that relation across twelve decades in wavelength.
The $68$~GHz, $1.5$~km design of \citetalias{gl15} lies on the $\vmax=100\kms$
locus; the same mission executed at $1\um$ needs a $23$~m aperture. Two horizontal
lines mark the envelopes we adopt in \S\ref{sec:beamform}: $10$~km at
$\sim20\,{\rm kW\,m^{-2}}$, and $40$~km, where the
emitters sit near the published $10^{3}\,{\rm W\,m^{-2}}$ loading. This suggests that
$20$--$100$~m optical apertures, within reach of existing terrestrial
engineering as collecting areas, are the Fresnel-matched counterparts of the
\citetalias{gl15} microwave array, but that they cannot radiate $1.5$~TW.

The beam power required to drive a perfectly reflecting sail is
$\PA=\ms \amax c/2$, i.e.\ $1.5$~TW for a tonne at $1$~gee, unchanged from
\citetalias{gl15} and roughly $10$\% of current worldwide consumption. The
diffraction scale of that aperture is
\begin{equation}
\thA = \frac{1.22\lambda}{\DA}
      = 13.9\,{\rm nrad}\,\lambda_{1.06}^{1/2}\,a_{1}^{1/2}\,v_{400}^{-1}.
\label{eq:thA}
\end{equation}
At the fiducials that is $2.9$~mas, and the corresponding beam solid angle is
$\Omega_{\rm A}=\pi\thA^{2}$. With the \citetalias{gl15} ratio $\Ds\simeq5\DA$ the
sail covers the beam core at burnout.

The $93$~m aperture that Fresnel matching wants must still radiate
$1.5$~TW, so the intensity leaving the aperture is
$\PA/A_{\rm A}=2.2\times10^{8}\,{\rm W\,m^{-2}}$, five orders above published
directed-energy loadings, a level that greatly exceeds the physical
capabilities of any known optical beam design.

Four ways out of that loading present themselves: (i) Oversizing the aperture until $I_{\rm A}$ is
tolerable puts $\dF$ beyond an Earth--Mars burnout, so the sail stays in the near
field. (ii) Lowering $\amax$ until $I_{\rm A}\sim10^{3}\,{\rm W\,m^{-2}}$ requires
$\amax\sim2\times10^{-3}$~gee, pushes $\DA$ back to $2$~km and $\dF$ to $25$~AU,
and again leaves a planetary transfer inside the near field. (iii) Sparse fill
runs the wrong way. At fixed $\PA$ and fixed envelope $\DA$, reducing the fill
factor $\phi$ shrinks the emitting area to $\phi A_{\rm A}$ and raises the
intensity on the emitters to $I_{\rm A}/\phi$, which at $\phi=10^{-5}$ is
$2\times10^{13}\,{\rm W\,m^{-2}}$, ten orders of magnitude above the
\citet{lubin14} value rather than comparable to it. (iv) Holding the emitters at
$10^{3}\,{\rm W\,m^{-2}}$ instead demands $1.5\times10^{9}\,{\rm m^{2}}$ of
emitting area, and at $\phi=10^{-5}$ an envelope of $1.4\times10^{4}$~km, for
which $\dF=D_{\rm env}^{2}/\lambda$ is $10^{9}$~AU and the sail never
leaves the near field.

Thus, an unfocused tube misses the sail however it is filled.
What remains is to spread $\PA$ over the area the loading demands and to apply
phase curvature, so that the tube focuses on the $465$~m sail instead of
illuminating a volume around it.

\subsection{The physical solution: a beam-formed emitter} \label{sec:beamform}

The intensity problem is an area problem. \citet{benford13} costed the
reusable launcher as source plus aperture, $C_{C}=aA+c_{P}P$, and
minimized that capital at fixed sail speed by cutting the beam when the
far-field spot first exceeded the sail; at the minimum the two terms
are equal, so $A=(c_{P}/a)P$. Two things in the focused optical machine
break that algebra. The mission power is pinned at $\PA=1.5$~TW by a
tonne at a gee, independent of $\lambda$ and $\DA$, and focusing puts
the sail in the near field of a large envelope, so that envelope can
still fill a $465$~m sail at $0.055$~AU. What remains of
$C_{C}=aA+c_{P}P$ at fixed power is an envelope that shrinks until
something else stops you; there is no interior cost minimum. The stop
we adopt is emitter loading. A free-flying orbital array, or an array
laid on an airless surface such as the Moon, has no atmosphere and so
no thermal blooming; the corner near $\sim20\,{\rm kW\,m^{-2}}$ is then
the loading the emitters themselves will bear, and at $1.5$~TW that is
$\DA\simeq10$~km ($I_{\rm A}=19\,{\rm kW\,m^{-2}}$), twenty times the
published directed-energy loading. Blooming at an
Earth-like ground site is the same number in a different guise. The published
loading of \citet{lubin14,lubin16} is a technological floor,
$I_{\rm A}=10^{3}\,{\rm W\,m^{-2}}$, not a Benford output; holding the
emitters there requires $\DA=43$~km, which we round to $40$~km
($I_{\rm A}=1.2\,{\rm kW\,m^{-2}}$). The optical coefficients of
\citet{parkin18} ($a=500$--$10^{4}\,{\rm \$\,m^{-2}}$,
$c_{P}=0.01$--$1\,{\rm \$\,W^{-1}}$) span $c_{P}/a$ from $10^{-6}$ to
$2\times10^{-3}$, i.e.\ $\DA=1.4$--$62$~km if one still wrote
$A=(c_{P}/a)\PA$; the $6$--$16$~km band is a diagonal slice through
that rectangle, not a selection. We treat $10$~km and $40$~km as two
emitter-loading choices that bracket the plausible range, the first
pushing the emitters and the second comfortable but aperture-heavy.

An unfocused beam of either diameter is a $10$~km or $40$~km tube, and a
$465$~m sail intercepts a fraction $(\Ds/\DA)^{2}=2.2\times10^{-3}$ or
$1.4\times10^{-4}$ of it, far short of what $1$~gee requires.
This suggests building that envelope as a phased array: $N$ small
emitters (tiles) laid on a regular lattice of pitch $p$ (the
distance between tiles), each able to take an independent delay.
A uniform extra delay of one tile, a piston of physical length
$\sigma_{z}$, is a phase $\sigma_{\phi}=2\pi\sigma_{z}/\lambda$ on
that tile alone. A linear gradient of those phases across the
aperture steers the interference peak; a quadratic curvature, the
thin-lens phase of focal length $d$, focuses it, concentrating the
tube onto the sail and shapes it to a $\Ds$-diameter flat-top. We take
$\Ds$ from the sail's thermal tolerance: spreading $1.5$~TW over
$8.7\times10^{6}\,{\rm W\,m^{-2}}$ gives $465$~m and $T\simeq300$~K at the
quoted absorptivity, coincidentally $5\DA$ of the Fresnel-matched $93$~m
aperture that cannot radiate this power. An unshaped diffraction-limited
focus at $\db$ would instead put $1.5$~TW into an Airy core of
$1.22\lambda \db/\DA$, $26$~cm across at $40$~km and $1.1$~m at
$10$~km, at $7\times10^{12}\,{\rm W\,m^{-2}}$ and
$4\times10^{11}\,{\rm W\,m^{-2}}$ respectively, which no sail
survives. In that unshaped geometry the sail edge sits $880$ spot radii
out at $40$~km and $220$ at $10$~km, and the missed fraction is
$f_{\rm miss}=1.9\times10^{-4}$ ($7.5\times10^{-4}$). That Airy tail is a floor on core spill,
not the lock of Table~\ref{tab:surveys}: a synthesized flat-top has edge
spill set by the achievable roll-off. The survey halo is set by piston
statistics, not by that core.

The far field of a tiled array is the product of two diffraction
patterns, the array factor (the interference of the $N$ tiles treated
as points, an interference peak whose width is set by the full envelope,
$\lambda/\DA$) and the element pattern (what one tile radiates by
itself, $\theta_{\rm el}=1.22\lambda/p$, set by the tile size and
independent of how the tiles are phased). A regular lattice also
produces grating lobes, extra copies of the array-factor peak at
multiples of $\lambda/p$, the Fourier comb of the grid. Locked phases
put the central peak on the sail. Uncorrelated piston does not
steer that peak: it reduces the fraction of power that remains in
the locked core, the Strehl $S=e^{-\sigma_{\phi}^{2}}$, exact for
Gaussian phase statistics rather than the Mar\'echal approximation, and dumps the
residual $1-S$ into the element pattern, a halo of width
$\theta_{\rm el}$. The element pattern is an envelope: the array factor can
redistribute power within it but cannot synthesize a footprint wider
than $\theta_{\rm el}$. Covering a $\Ds$ sail therefore requires
$1.22\lambda \db/p\gtrsim\Ds/2$, i.e.\ $p\lesssim45$~m at the
fiducials. We adopt $p=40$~m, a round pitch just inside that thermal
cap; encircled energy and $f_{\rm leak}$ barely move, while
$\theta_{\rm el}$ and the umbra in units of $\theta_{\rm el}$ do.
Violate it and the footprint shrinks below the sail: $N=100$ on the
$40$~km envelope is $p\simeq3.5$~km ($0.89$~km on $10$~km) and a
$\sim3$~m spot at burnout, $\sim2\times10^{11}\,{\rm W\,m^{-2}}$, four
orders above the $8.7\times10^{6}\,{\rm W\,m^{-2}}$ flat-top and two
above the dust-seeded hot-spot failure of \citet{jaffe23}. The sail
does not survive. Tiles individually steered, or laid on a curved
surface so that different tiles illuminate different parts of the
sail, would beat the envelope only by giving up coherent gain; we do
not take that architecture. Smaller pitches leak more at the same
lock and at fixed $\Ds$, so the $\lambda/10$ value $f_{\rm leak}=0.054$ is a floor over
thermally viable pitches. Metre-class modules, the published-concept
end of that range, leak a third of $\PA$ because the sail encircles
only $0.18$~percent of a halo of first-zero radius $10.6$~km; that is a
visit-scale $m\sim23$ bump, and Table~\ref{tab:surveys} does not
constrain it. Filling $10$~km at $p=40$~m gives
$N=(\pi/4)(\DA/p)^{2}=4.9\times10^{4}$ tiles, and $40$~km gives
$N=7.9\times10^{5}$.

For an intercept-matched pitch $p=2.44\lambda \db/\Ds$, the element
scale is $\theta_{\rm el}=\Ds/(2\db)$ exactly, the sail's angular radius
as seen from the array, and $\tau_{\rm el}=\Ds/(2 v_{\perp})$, with
$\lambda$, $p$ and $\DA$ all cancelling. The first-ring
isotropic-equivalent luminosity scales as
$\Liso\propto I_{\rm pk}(1-S)\PA (\db/\Ds)^{2}$, again independent of
$\lambda$ and $\DA$. The observable set $\{\theta_{\rm el},\taueff,\Liso\}$
depends only on $(\PA,\Ds,\db,\sigma_{z}/\lambda,v_{\perp})$, and either
envelope drops out of the survey prediction. We use $p=40$~m
rather than $45.4$~m, so this cancellation is approximate, but it is
why $\taueff$ is identical in every row of Table~\ref{tab:surveys} and
why the row-to-row $m_{\rm pk}$ differences are instrumental, and why the
$10$~km and $40$~km arrays overlie in Figures~\ref{fig:design}b
and~\ref{fig:limits}. A higher
tolerable loading $I_{\rm A}$ shrinks $\DA\propto I_{\rm A}^{-1/2}$
without changing the halo observables of this section; $10^{3}\,{\rm W\,m^{-2}}$
is a technological choice, not a physical one. Smaller payloads, longer
burns at lower $\amax$ with the array still focused, and staged launches
leave that loading unchanged.

\begin{figure*}
\centering
\includegraphics[width=0.918\linewidth]{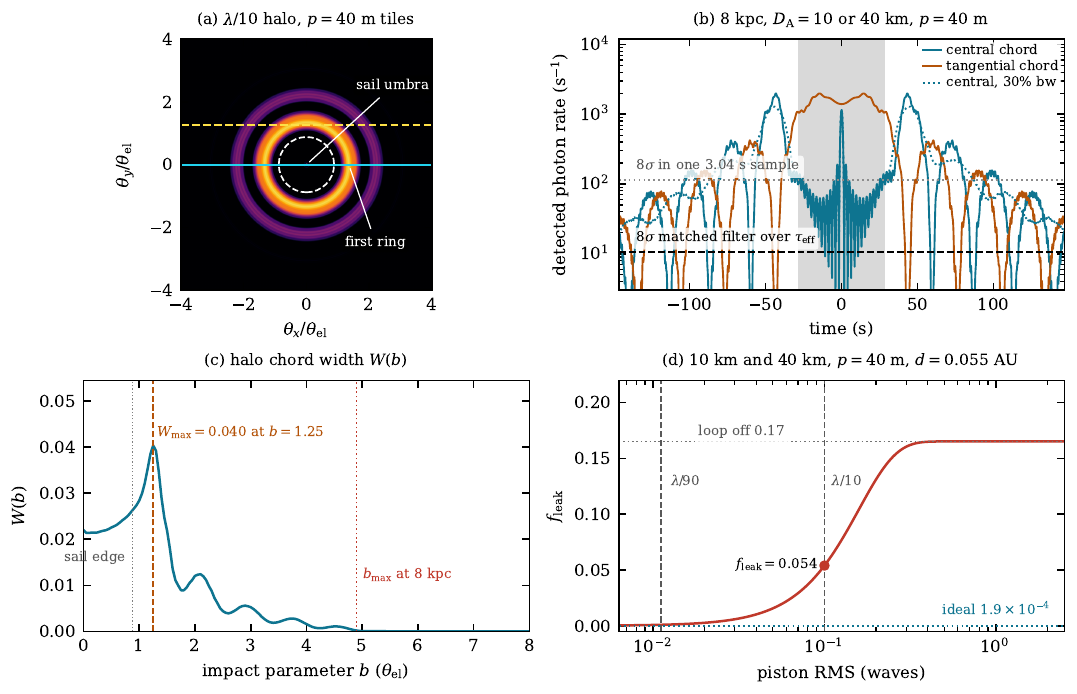}
\caption{(a) Leaked far-field of a $10$~km or $40$~km array of $40$~m tiles focused on
the $465$~m sail at $\db=0.055$~AU, each tile delayed independently by
an RMS piston $\sigma_{z}=\lambda/10$ (a uniform extra path on that
tile). Axes are in units of the element diffraction scale
$\theta_{\rm el}=1.22\lambda/p$. The pattern is the Fraunhofer transform of
one tile's sail-plane Airy truncated by the sail, smoothing scale
$\lambda/\Ds\simeq2.3$~nrad ($0.07\,\theta_{\rm el}$). The dashed circle is the geometric sail edge
($0.88\,\theta_{\rm el}$). $I$ is in units of the unocculted tile peak, so the
plotted pattern carries scattered power $(1-S)\PA$ and the first-ring
$\Liso=I_{\rm pk}\,4\pi(1-S)\PA A_{\rm el}/\lambda^{2}$. An ideal lock ($S=1$) leaves this panel
dark. Lines mark the central ($b=0$) and tangential ($b=1.25$) chords.
(b) Light curves at $8$~kpc in F146. The shaded interval is the geometric
umbra. The dashed line is the peak rate at which a matched filter over
$\taueff$ reaches $8\sigma$, the dotted line the harder cut a single $3.04$~s
sample would need; the central chord is a pair of first-ring peaks $87$~s
apart at ${\rm S/N}\simeq810$, the tangential chord a single peak at
${\rm S/N}\simeq1.5\times10^{3}$. The dotted central curve is a $30$~percent band, which
fills the dip from $0.58$ of the peak to $0.90$. (c) Chord equivalent width of that
halo, Equation~(\ref{eq:W}), $b$ in units of $\theta_{\rm el}$. $W$ peaks at
$0.040$ at $b=1.25$; $b_{\rm max}\simeq4.9$ at $8$~kpc. (d) Missed power versus RMS tile
delay in waves of $\lambda$. The dotted line is the unshaped-focus
$f_{\rm miss}=1.9\times10^{-4}$ at $40$~km ($7.5\times10^{-4}$ at $10$~km); the $\lambda/10$ point is
$f_{\rm leak}=0.054$. Unlocking the loop saturates at $0.17$ because
the tile Airy still hits the sail. The survey halo is set by piston
statistics, not by that core spill.}
\label{fig:beamform}
\end{figure*}

We take a residual piston $\sigma_{z}=\lambda/10$ ($100$~nm), the leftover
error of a feedback loop that holds each tile rather than the
laboratory lock of \citet{fsaifes20}, and find $S=0.67$ and
$f_{\rm leak}=0.054$, a floor at this lock and fixed $\Ds$, $80$~GW missing the sail
(Figure~\ref{fig:beamform}d). The sail then
encircles $0.83$ of the element Airy; unlocking the loop cannot dump
the whole beam, and $f_{\rm leak}$ saturates at $0.17$. At
$\lambda/90$ the extra halo is $0.08$~percent. A $4$~cm sideways
placement of the outer tiles is another $\lambda/10$ of path to the
sail and raises $f_{\rm leak}$ to $0.074$. A common pointing error of
$\sim3$~nrad slides the locked core by $25$~m, $\sim10$~percent of the
sail radius, and leaves $f_{\rm leak}$ unchanged; $30$~nrad would dump
the core off the sail entirely ($245$~m at $\db$). Pointing at the
element scale is therefore comparable to the $\lambda/10$ piston lock.
We model only that uncorrelated piston, which we treat as the single
free parameter. Intra-tile figure error on a $40$~m aperture would
broaden $\theta_{\rm el}$ and cut the fluence as $\theta^{-1}$; correlated
(common-mode) piston scatters into the core rather than the halo and
reduces $f_{\rm leak}$. Both omitted terms push the same way on the
observable, so the predicted flash is an upper envelope on this lock.

Structure
in the leak, grating lobes and a speckle of $N$ random phasors, appears
when the pitch is small enough that those features miss the sail
($p\lesssim40$~m) but $N$ is still modest; that is a laboratory tiled
pupil \citep{fsaifes20}, not this intensity-limited array.

The leaked field of one tile is the Fraunhofer transform of the
tile's sail-plane Airy truncated by the sail disk \citep{goodman17}, smoothing scale
$\lambda/\Ds\simeq2.3$~nrad ($0.07\,\theta_{\rm el}$). $I$ is in units of
the unocculted tile peak, which carries scattered power $(1-S)\PA$, so
the first-ring $\Liso=I_{\rm pk}\,4\pi(1-S)\PA A_{\rm el}/\lambda^{2}=0.38\,L_{\odot}$
at $I_{\rm pk}=0.022$. A central observer sees two first-ring peaks
$\sim30$~s across and $87$~s apart (the intensity-ring peak at $b\simeq1.33$,
not the $W$ peak), with the on-axis intensity $0.58$ of the peak rather
than a dark umbra; a $30$~percent band fills that dip to $0.90$ of the
peak, so the doublet contrast is a narrowband discriminant
(Figure~\ref{fig:beamform}b). The geometric umbra is
$0.88\,\theta_{\rm el}$ against $b_{\rm max}\simeq4.9$, so about one
detection in five is such a pair, and the rest are a single
first-ring transit. The tangential chord at $b=1.25$, where $W$
peaks at $0.040$, is that single peak, $m_{\rm AB}\simeq19.4$ in F146 at
$8$~kpc, which a matched filter over $\taueff$ records at
${\rm S/N}\simeq1.5\times10^{3}$; the central doublet is the same first-ring
fluence at ${\rm S/N}\simeq810$
(Figure~\ref{fig:beamform}a--c), and that is the halo of
Table~\ref{tab:surveys} and Figures~\ref{fig:design}b and~\ref{fig:limits}.
We adopt the $10$~km and $40$~km focused arrays as the practical
optical designs; a Fresnel-matched optical aperture cannot radiate $1.5$~TW,
and the halo of this section is the leak a Galactic survey would catch.

\subsection[From 10 GW to 10 TW]{From $10$~GW to $10$~TW} \label{sec:power}

The leaked fraction is set by the lock and the intercept, not by the
radiated power. What $\PA$ sets is the payload. Varying the mass at
$1$~gee, $\PA=\ms\amax c/2$ runs from $10$~GW at $7$~kg to $10$~TW at
$7$~tonnes, with the tonne of \citetalias{gl15} at $1.5$~TW.
High-power sites on Earth have entered the lower decade of that range.
Data-center campuses built to train large models are now specified at
gigawatts \citep{iea25}, and the \emph{Stargate} program has committed
$10$~GW of dedicated capacity in the United States \citep{openai26}, of
order a large nuclear plant and a tenth of the $0.1$~TW Saturn~V that
\citetalias{gl15} compared to a tonne. A $10$~GW beamer is a megaproject,
but it is a megaproject of a kind already being assembled, if for
computation rather than propulsion. It accelerates $7$~kg at $1$~gee.
The tonne still requires $1.5$~TW, $150$ such plants, and remains
$\sim10$\% of worldwide consumption.

\begin{figure*}
\centering
\includegraphics[width=0.918\linewidth]{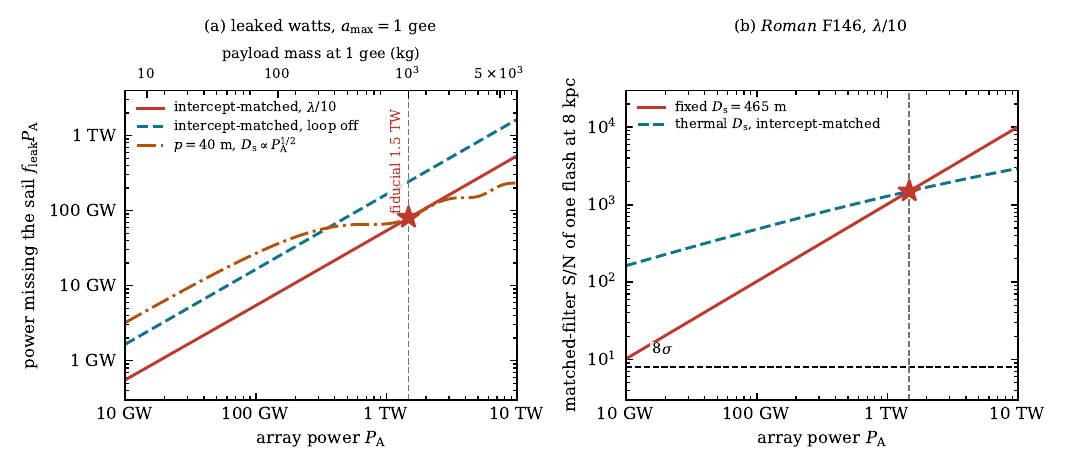}
\caption{(a) Power missing the sail as a function of array power, from
$10$~GW ($7$~kg at $1$~gee) to $10$~TW ($7$~tonnes). The solid and dashed
lines hold the intercept match, so $f_{\rm leak}$ is constant and the
leaked power tracks $\PA$; unlocking the loop multiplies the leak by
three. The dash-dotted line freezes $p=40$~m and grows the sail with the
thermal loading, $\Ds\propto\PA^{1/2}$: a small sail misses the halo and
a large sail swallows it. The star is the fiducial tonne at $1.5$~TW.
The upper axis converts $\PA$ to payload mass at $1$~gee.
(b) Matched-filter S/N of one $\lambda/10$ first-ring flash at $8$~kpc
in F146. Holding $\Ds=465$~m, $\Liso$ tracks $\PA$ and the $8\sigma$
floor is $8$~GW. Growing the sail and re-matching the pitch, $\Liso$ is
independent of $\PA$ ($m_{\rm AB}\simeq19.4$) and the S/N still rises
because $\taueff\propto\Ds$. The two curves meet at the fiducial.}
\label{fig:power}
\end{figure*}

At the $\lambda/10$ intercept-matched lock that is $0.54$~GW leaking at
the low end and $540$~GW at the high end, a straight line through the
$80$~GW of a tonne (Figure~\ref{fig:power}a). Unlocking the loop
multiplies the leak by three ($f_{\rm leak}=0.17$). If instead the
$40$~m pitch is frozen and the sail is grown with the thermal loading,
$\Ds\propto\PA^{1/2}$, a $10$~GW payload is a $38$~m sail that encircles
only $0.019$ of the element Airy and leaks $0.32$ of $\PA$, while a
$10$~TW cargo is a $1.2$~km sail that swallows the halo down to
$f_{\rm leak}=0.023$. The two $\lambda/10$ curves meet at the fiducial,
where $p=40$~m is the intercept match.

What a survey records is not that wattage but the first-ring $\Liso$.
Holding $\Ds=465$~m and turning the array up or down, $\Liso$ tracks
$\PA$ and \emph{Roman}'s $8\sigma$ floor at $8$~kpc sits at $8$~GW, so
the left edge of Figure~\ref{fig:power}b is where a tonne-class sail
driven at reduced power drops through threshold. Growing the sail and
re-matching the pitch, the halo solid angle grows with the power,
$\Liso$ is independent of $\PA$, and the flash stays at
$m_{\rm AB}\simeq19.4$; the matched-filter S/N still rises, from $160$
at $10$~GW to $2.9\times10^{3}$ at $10$~TW, because $\taueff\propto\Ds$.
At fixed loading the envelope scales as $\DA\propto\PA^{1/2}$,
$0.8$--$26$~km at $19\,{\rm kW\,m^{-2}}$ and $3.6$--$113$~km at
$10^{3}\,{\rm W\,m^{-2}}$ over the same range, and the halo observables
of \S\ref{sec:beamform} do not depend on that envelope. A thermally sized
$10$~GW sail is already $m_{\rm AB}\simeq19.4$ and
${\rm S/N}\simeq160$ at $8$~kpc, a Galactic flash; we retain
$\PA=1.5$~TW as the fiducial because it is a tonne at a gee.

\subsection{Survival of the sail and its reflection} \label{sec:refl}

Naively one might spread $\PA$ over the full sail and find
$I_{\rm s}=4\PA/\pi\Ds^{2}=8.7\times10^{6}\,{\rm W\,m^{-2}}$ and $T\simeq300$~K.
For the unfocused $93$~m aperture the beam cross-section is $\sim\DA$, not $\Ds=5\DA$, for
most of the burn; the sail is oversized to catch the beam only near $\dF$. The
illuminated spot therefore carries $2.2\times10^{8}\,{\rm W\,m^{-2}}$ early and
$\sim5\times10^{7}\,{\rm W\,m^{-2}}$ at burnout. With absorptivity
$\alpha=10^{-4}$ at $1\um$ and two-sided radiation at infrared emissivity of order
unity, the sail equilibrates at $T\simeq660$~K at the start and $T\simeq460$~K at
burnout. The focused $10$~km and $40$~km arrays of \S\ref{sec:beamform} shape a
$\Ds$-diameter flat-top onto the sail at every $d$, so that
$8.7\times10^{6}\,{\rm W\,m^{-2}}$ and $T\simeq300$~K hold for the whole burn
rather than only near $\dF$; the $\lambda/10$ lock leaves $S=0.67$ of $\PA$ in
that core and the sail still encircles $0.83$ of the element Airy, and the
on-axis intensity at burnout is $\simeq1.4\times10^{7}\,{\rm W\,m^{-2}}$
($T\simeq330$~K). A factor-of-sixteen peak on that flat-top,
$(600/300)^{4}$, would still be only $T\simeq600$~K. The $300$--$660$~K range is a
factor of two to five below the $\sim1600$~K vacuum-decomposition threshold of
silicon nitride \citep{gao24}, and sits at the cool end of the $500$--$1500$~K
window already assumed for Starshot thermal design \citep{atwater18,lin25}. The failure
mode at $1\,{\rm GW\,m^{-2}}$ is a dust-seeded hot spot
\citep{jaffe23}, two orders above the focused flat-top and a factor of a few
above the unfocused spot. Lower accelerations relieve the thermal
load (the sail intensity scales as $\amax$, the aperture intensity as $\amax^{2}$)
at the cost of a larger array. The fiducial
sail is $465$~m across, area $1.7\times10^{5}\,{\rm m^{2}}$, at a total system
mass of $10^{3}$~kg. A $0.2$--$1\,{\rm g\,m^{-2}}$ film in the Starshot range
\citep{atwater18,parkin18} is then $34$--$170$~kg of sail, leaving
$830$--$966$~kg for payload, structure and thermal control; the
$6\,{\rm g\,m^{-2}}$ figure is that total under $1$~gee, aggressive even by
Starshot standards.

The sail is a near-perfect mirror and returns essentially the intercepted
power, $\simeq0.95\,\PA$ at the $\lambda/10$ lock. A flat retro-reflector
sends that return onto the array at $\sim8\times10^{6}\,{\rm W\,m^{-2}}$,
some $400$ times the $10$~km array's own emission loading, whether the
array is a free-flyer or is laid on an airless surface
(\S\ref{sec:beamform}). Leaving the return on the array still faces
$400$ times the emission loading in orbit or on an airless world, which
is intolerable unless the return is steered off the aperture. Deflection
is a requirement, not an optional geometry. Two ways remain to keep the
return off the array: metasurface beam-riding sails
\citep{gao24} deflect rather than retro-reflect, sending the return
along a third axis, or that deflected beam is aimed at a recapture
site, a solar collector that recovers a fraction of $\PA$. Recapture
points at a fixed collector and paints no strip on the sky. Deflection
alone is a fixed angle off the array--sail line, so the strip sweeps
with that line through the full burn, and
$\Liso^{\rm (refl)}=4\PA/\theta_{\rm refl}^{2}$ is independent of $d$,
so $\Delta\phi_{\rm eff}$ is the whole $\sim1.1$~rad rather than
$5.1\times10^{-5}$. The deflected strip is then a couple of hundred
times larger in solid angle than the leak despite being $\sim14$ times
narrower ($\theta\sim\lambda/\Ds$,
$\Liso^{\rm (refl)}\simeq2\times10^{3}\,L_{\odot}$). What kills it is
duration: $\tau=\theta_{\rm refl}\,\db/v_{\perp}\simeq2.4$~s at burnout
and shorter earlier, a single-read spike, exactly the regime where the
persistence cut of \S\ref{sec:backgrounds} is unavailable and the
$10^{7}$ snowballs come back. We take deflection off the array, with or
without recapture.
If the sail still retro-reflects, a sag $\sigma$ (a cone or spherical cap, the
geometry assumed for beam-riding stability before those designs)
spreads that return over $\theta_{\rm refl}\simeq8\sigma/\Ds$.
Attitude wander $\delta$ during the arrival burn smears the same beam by $2\delta$,
and holding a $465$~m membrane to microradians for eleven hours is the binding
requirement, tighter than figure at $\sigma/\Ds$. The effective width is the
quadrature of those two terms. Then $\Liso^{\rm (refl)}=4\PA/\theta_{\rm refl}^{2}$
and the duration is $\theta_{\rm refl}/\dot\phi$.

Those two limits do not overlap as a second flash. A flat sail is
$\sim50$ times narrower than the halo strip half-width
$b_{\rm max}\theta_{\rm el}\simeq160$~nrad at the bulge, so an observer who sees the launch
leak has probability $\sim2\times10^{-2}$ of also lying in the reflection, and
that assumes the reflection axis coincides with the leak axis. A surface sag
that produces $\theta_{\rm refl}=1^{\prime\prime}$ ($\sigma\simeq0.3$~mm) covers every leak-observer and is detected at $8$~kpc
($34\sigma$ in a matched filter), but it sweeps past in
$\theta_{\rm refl}/\dot\phi\simeq5\times10^{3}$~s, an hour-scale brightening, not
a second entry in a tens-of-seconds ramp search. The region of $(\sigma,\delta)$ in which
the reflection is both geometrically probable and a tens-of-seconds flash is
empty: a $10$--$100$~s reflection has overlap $0.01$--$0.1$ with the leak strip,
and covering that strip forces the duration to hours. The reflection is not a
second Galactic yield channel, and it is not a pair of flashes. The pair that
does survive is internal to the leak, the two first-ring crossings on either
side of the sail's shadow (\S\ref{sec:signal}).

\section{The observable event} \label{sec:signal}

\subsection{Duration}

Because the sail must be tracked, the beam sweeps, and \citetalias{gl15} found of
order a radian of azimuthal motion during a three-hour burn, with the sweep fastest
early (when the sail is near the array) and slowest at burnout, where the
isotropic-equivalent luminosity peaks (their Figure~3). The 2015 trajectory is
specified completely: the sail starts in low orbit about the origin world, and the
array beams along the world--sail line, so the force is radial and the specific
angular momentum $L=v_{\rm LEO}R_{\rm LEO}$ is conserved about that world. We take
a $400$~km circular orbit, $R_{\rm LEO}=6.77\times10^{6}$~m and
$v_{\rm LEO}=7.67\kms$, so $L=5.20\times10^{10}\,{\rm m^{2}\,s^{-1}}$. Integrating
the same three-body trajectory as \citetalias{gl15} gives a full-burn sweep
$\Delta\phi\simeq1.1$~rad and a residual transverse speed at burnout of
$6.4\,{\rm m\,s^{-1}}$ from $L/\db$ alone. The start-from-rest integral
$\Delta\phi=(4L/3\amax^{2})(t_{1}^{-3}-t_{\rm F}^{-3})$ with
$d=\tfrac12\amax t^{2}$ from $d_{1}=R_{\rm LEO}$ gives $0.44$~rad; the factor
$2.5$ is the initial orbital velocity and the non-zero starting radius. The Sun
and the destination contribute a few meters per second at $\db$, a
$30$--$50$ percent correction to that residual; we add $5\,{\rm m\,s^{-1}}$ in
quadrature and use $v_{\perp}(\db)=8.1\,{\rm m\,s^{-1}}$.

In closed form, the residual transverse speed at the end of the burn is
\begin{equation}
v_{\perp}(\db)=\frac{L}{\db}=\frac{v_{\rm LEO}R_{\rm LEO}}{\vmax^{2}/(2\amax)},
\label{eq:vperp}
\end{equation}
and the time for a diffraction scale $\theta$ to sweep past a fixed observer is
$\tau=\theta/\dot\phi=\theta\,\db/v_{\perp}(\db)$. For the
\citetalias{gl15} microwave design that recovers $v_{\perp}=0.10\kms$ and a
duration $\thA \db/v_{\perp}=18$~s. For the favored halo the scale is the
element pattern, $\theta_{\rm el}=1.22\lambda/p=32$~nrad at $p=40$~m, and
$\tau_{\rm el}=\theta_{\rm el}/\dot\phi=33$~s.

\begin{figure*}
\centering
\includegraphics[width=0.986\linewidth]{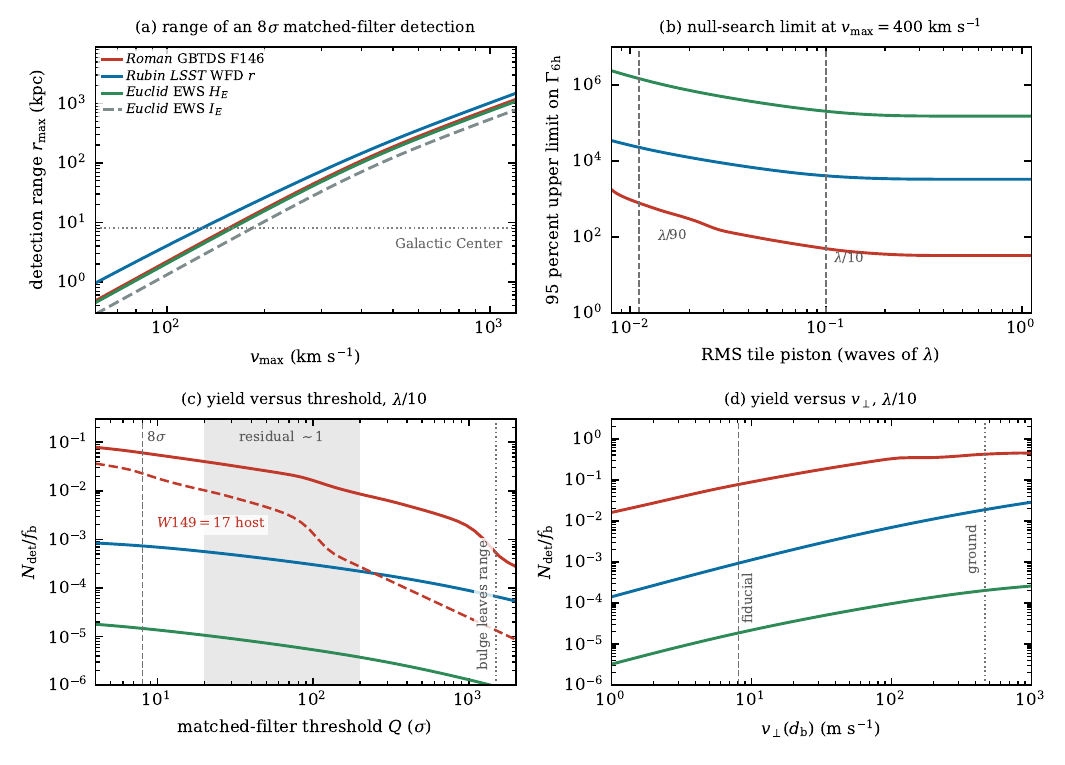}
\caption{(a) Range of an $8\sigma$ matched-filter detection of the
$\lambda/10$ halo as a function of
transit speed, for a $10^{3}$~kg payload at $1$~gee on a $10$~km or $40$~km array of
intercept-matched tiles. (b) The $95$ percent upper
limit on $\GamL$ that a null search would imply, as a function of
RMS tile piston, from Equation~(\ref{eq:yield}) at
$\vmax=400\kms$. Vertical lines mark a laboratory lock ($\lambda/90$) and the
fiducial floor ($\lambda/10$). A $\lambda/90$ lock does not reach the bulge.
Curves in (a) terminate where the signal
falls below threshold at the distance of the bulk of the stellar population.
\emph{Euclid} does not appear as a useful limit in (b). (c) Yield as a function
of the matched-filter threshold. Because $r_{\rm max}$ already exceeds the
Galaxy at $8\sigma$ on a faint host, a harder cut removes strip width and
detectable sweep rather than stars, until $r_{\rm max}$ falls to the distance of
the bulge near $Q\simeq1.5\times10^{3}$. The dashed curve is \emph{Roman} with a
$W149=17$ host (host shot noise, filling factor unity, full column), an upper
bound on a giant-host search. The shaded band is the decade in $Q$ at which a
$10^{-4}$--$10^{-6}$ survivor fraction of the duration-cut sample of
\S\ref{sec:backgrounds} would leave of order one event. (d) Yield versus
transverse speed at burnout. Colors are as in (a).}
\label{fig:limits}
\end{figure*}

These durations assume an array whose transverse velocity $u$ relative to the
origin world's center is small compared with $v_{\perp}(\db)$. An equatorial
ground site has $u\simeq465\,{\rm m\,s^{-1}}$, a geostationary platform
$u\simeq3.1\kms$, and a low-orbit array $u\simeq7.8\kms$, which would shorten
$\tau_{\rm el}$ by factors of $58$, $390$ and $10^{3}$. The fiducial burn lasts
$\vmax/\amax=11.3$~hours, longer than a target stays above the horizon at a
single ground site, so a ground array would also have to hand off between sites
and the sweep would not be the smooth $L/d^{2}$ used here. This suggests a
free-flying array co-orbital with the origin world, the architecture assumed by
most large-aperture directed-energy concepts \citep{lubin16,parkin18}. Holding
that array to $u\ll8\,{\rm m\,s^{-1}}$ over an $11$~hour burn is a
station-keeping requirement we do not solve; we instead recompute $N_{\rm det}$
and $\taueff$ as functions of $v_{\perp}(\db)$ from $1$ to
$10^{3}\,{\rm m\,s^{-1}}$ (Figure~\ref{fig:limits}d). Because $\taueff$ and
$\Delta\phi_{\rm eff}$ trade, $N_{\rm det}/\fb$ runs from $0.016$ at
$1\,{\rm m\,s^{-1}}$ to $0.33$ at $100\,{\rm m\,s^{-1}}$, of order the
Table~\ref{tab:surveys} value near the fiducial $8\,{\rm m\,s^{-1}}$. A ground
site at $465\,{\rm m\,s^{-1}}$ is a $1$~s flash: $r_{\rm max}=17$~kpc still
exceeds the bulge (${\rm S/N}=35$ at $8$~kpc) and the integrated yield is
$0.42\,\fb$, but the event is a single-read spike rather than a $60$~s ramp,
and a ground array must still hand off between sites.

A distant observer whose line of sight passes at impact parameter $b$ from the
pattern center records a chord through the leaked far-field. Integrating along
that chord gives the event fluence, which we write as
$\taueff(b)=(W(b)/I_{\rm pk})\,\tau_{\rm el}$ with
\begin{equation}
W(b)=\int I(x)\,du,\qquad x=1.22\pi\sqrt{u^{2}+b^{2}},
\label{eq:W}
\end{equation}
with $u$ and $b$ in units of $\theta_{\rm el}$, so $x=\pi p\theta/\lambda$ is
the standard Airy argument \citep{goodman17}, and $I$ the leaked far-field of the truncated
tile Airy in units of the unocculted tile peak. $W$ peaks at $0.040$ for
$b=1.25$ (Figure~\ref{fig:beamform}c), the leaked first-ring peak is
$I_{\rm pk}=0.022$, and $\taueff\simeq60$~s. At the bulge
$b_{\rm max}\simeq4.9$ in units of
$\theta_{\rm el}$, so the events that make up the \emph{Roman} yield are chords
through the first ring and its inner Airy wings, not a $b^{-2}$ tail out to tens
of $\thA$.

The durations above are the ones at burnout. Earlier in the burn the sail covers more
of the element Airy, $f_{\rm halo}$ is smaller, and the first ring has not yet
cleared the occulter. For a bulge star the end-of-burn halo fluence exceeds the
\emph{Roman} threshold by a factor $(r_{\rm max}/8\,{\rm kpc})^{2}\simeq190$, which
with the computed $d$-scaling of the occulted element pattern corresponds to
$d_{\rm min}/\db\simeq0.31$ and $\Delta\phi_{\rm eff}=5.1\times10^{-5}$~rad. The
full $1.1$~rad of the \citetalias{gl15} trajectory is accumulated almost entirely in the first
minutes, when the leak is suppressed. Naively using the full-burn sweep
overestimates the geometric probability by four to five orders of magnitude. The
start-from-rest integral is adequate for this interval, since the detectable
window has $d_{1}/\db\gtrsim0.05$ and therefore $d_{1}\gg R_{\rm LEO}$, where it
agrees with the three-body integration. We use $\Delta\phi_{\rm eff}$ in the
yield.

\subsection{Appearance in an imaging survey}

Figure~\ref{fig:beamform} shows the halo and two representative light curves for a
system at $8$~kpc. At peak the flash reaches $m_{\rm AB}\simeq19.4$ in the F146
photon-rate zeropoint, a factor $\sim17$ above the \emph{Roman} $8\sigma$ single-read
line, and lasts $\taueff\simeq60$~s. Four properties follow, and together they
define the search.

\emph{(i) The event is short compared with a visit and long compared with a cosmic
ray \citep{rauscher25}.} The first-ring pair of Figure~\ref{fig:beamform}b is not resolved as a
spatial ring; the survey measures its fluence. On \emph{Roman},
$\taueff\simeq60$~s spans $\sim\!20$ reads of $3.04$~s, so that fluence is a short
ramp transient rather than a single-sample jump \citep{casertano22}. A central chord is two peaks of
duration $\sim30$~s separated by $87$~s, so \emph{Roman}'s ramp records
a doublet in one visit; a tangential chord is a single peak, and that single peak
is the typical detection, the umbra being missed off axis. \emph{Rubin}'s
$30$~s visits (Table~\ref{tab:surveys}; \citealt{ivezic19}) and \emph{Euclid}'s $87$~s NISP
integrations \citep{jahnke25} generally do not resolve that doublet; $2\times15$~s snap mode is a
recommended option (\S\ref{sec:rubin}), not the as-executed baseline. The search we advocate is a
matched filter over the flash window
(Equation~\ref{eq:snr}). For a fixed-fluence
transient the background noise is $\sqrt{B\taueff}$, independent of $t_{\rm samp}$;
read noise worsens as $t_{\rm samp}$ shrinks if reads are combined incoherently.
The argument for short sampling is instead localization of the flash
window, which matters when $\taueff$ is much shorter than the exposure (the
\emph{Euclid} VIS case, $t_{\rm samp}=566$~s; \citealt{cropper25}). Figure~\ref{fig:beamform}b shows both the matched-filter
threshold and the harder cut a single $3.04$~s sample would need; the ranking of
the three facilities is set by collecting area and by how well $t_{\rm samp}$
matches $\taueff$.

\emph{(ii) The flash is single-band, not necessarily a spectral line.} For the
focused array the relevant chromaticity is whether the focus and the tile lock
are true time delay, which is achromatic, or phase, for which
$\sigma_{\phi}=2\pi\sigma_{z}/\lambda$ varies by $\sim30$~percent across a
$30$~percent band \citep{fsaifes20}. Either way a $10$--$30$ percent fractional bandwidth still
lands all of the flash energy in one filter. Figure~\ref{fig:beamform} is
monochromatic at the Nd:YAG line
$\lambda=1.06\um$ \citep{lubin16,parkin18}; a $30$ percent band broadens $\theta_{\rm el}$ by that fraction
and fills the far-field dip from $0.58$ of the peak to $0.90$, so the doublet
contrast is a narrowband discriminant and does not restore a core. That is the basis of a color veto on a single flash:
astrophysical variability is achromatic or smoothly chromatic \citep{kowalski16}. An unresolved
feature on a stellar trace in NISP grism data at $R\sim450$ \citep{jahnke25} would require
$\Delta\lambda/\lambda\lesssim2\times10^{-3}$, two orders of magnitude tighter
than a $30$~percent band, and we do not assume it \citep{tellis17,zuckerman23}.

\emph{(iii) The source is unresolved and coincident with a star.} At $8$~kpc the
fiducial $\db=8.2\times10^{9}$~m ($0.055$~AU) subtends $7\,\mu$as. The flash is
therefore a perfect point source at the stellar position, which is what permits
the background rejection of \S\ref{sec:backgrounds}.

\emph{(iv) Launch and arrival leaks are antiparallel.} Arrays on both worlds
so that the cargo can be decelerated \citep{forward84} imply two beams. At launch, $A$ beams along
$A\to B$ and the leaked radiation continues toward $B$ and beyond. At arrival, $B$
beams head-on at the incoming sail, along $B\to A$, and its leak continues toward
$A$ and beyond. The two leaked strips are antiparallel, so an observer beyond $B$
sees the launch flash and is not illuminated at arrival; the geometric factor of
$2$ in Equation~(\ref{eq:pgeo}) counts those disjoint strips, one flash per
transfer per observer. The reflected beam restores a pair only in a thin
corner of sail figure and attitude (\S\ref{sec:refl}). The pair that does
survive is internal to one burn: a line of sight that clips the axis crosses
the first ring on both sides of the shadow, and \emph{Roman}'s ramp records two
peaks $87$~s apart (Figure~\ref{fig:beamform}b). The umbra is $0.88\,\theta_{\rm el}$ against
$b_{\rm max}\simeq4.9$, so about one detection in five is such a doublet, and
the rest are single first-ring transits.

A second flash from a later transfer is a much weaker prospect. One burn paints
a strip $\Delta\phi_{\rm eff}=5.1\times10^{-5}$~rad long and
$2b_{\rm max}\theta_{\rm el}=3.2\times10^{-7}$~rad wide, and the strip lies
along the line joining the two worlds, which turns at
$\sim\!10^{-7}\,{\rm rad\,s^{-1}}$ as they orbit and so moves
$\sim\!2\times10^{-3}$~rad between transfers six hours apart, some forty strip
lengths. Later transfers lay their strips elsewhere along the band that rotating
line sweeps, and an observer already in the band is caught again with
probability $\Delta\phi_{\rm eff}/2\pi=8\times10^{-6}$ per transfer. Over the
$157$ transfers a GBTDS star is watched through at one per $6$~h, the expected
number of repeats is $1\times10^{-3}$, and a repeat inside the survey would need
a launch every $\sim\!30$~s. A single flash is therefore a photometric
candidate; the typical detection is a single first-ring peak, the doublet is a
$\sim20$~percent subset whose morphology is the confirmation test, and the
statistical cut of \S\ref{sec:backgrounds} is what a single flash has to
survive. Because
the laser wavelength sets the intercept-matched pitch, the four rows of
Table~\ref{tab:surveys} are four distinct beamers on either envelope;
a \emph{Roman} candidate
is followed up in F146, not by asking \emph{Rubin} or \emph{Euclid} to catch
the same flash in $r$ or $H_E$.

\section{Detectability with \textit{Roman}, \textit{Rubin} and \textit{Euclid}} \label{sec:surveys}

\subsection{The right statistic} \label{sec:stat}

Point (i) above implies that the search should not be conducted on stacked light
curves. For the near-infrared detectors of \emph{Roman} and \emph{Euclid}/NISP,
exposures are sampled up the ramp and the pipelines already compute, for every
pixel and every read, the jump between successive samples in order to identify
cosmic rays \citep{casertano22}. A tens-of-seconds flash appears in this data product as a run of
consecutive jumps that is (a) confined to $\sim\!\taueff$, (b) distributed across
the pixels of the stellar point-spread function (PSF) in proportion to the PSF, and (c) centered on a
cataloged star. We therefore advocate a {\it PSF-matched ramp search}: fit each
ramp segment with the PSF at the position of each cataloged source and record the
amplitude and its $\chi^{2}$. For \emph{Rubin} the analog is a search of
individual $15$~s snaps or single visits against the deep template \citep{ivezic19}, retaining
PSF-consistent positive residuals. In each case the matched-filter significance
over the flash window is
\begin{equation}
\frac{S}{N}=\frac{S_{\rm peak}\taueff A_{\rm eff}\,E_{\gamma}^{-1}}
{\left[(\dot N_{\rm src}\!+\!\dot N_{\rm sky})\taueff+2\sigma_{\rm read}^{2}n_{\rm pix}\right]^{1/2}},
\label{eq:snr}
\end{equation}
where $E_{\gamma}=hc/\lambda$ is the photon energy, $\dot N_{\rm src}$ is the
host stellar photon rate in the aperture, and $A_{\rm eff}$ is the
effective collecting area (Table~\ref{tab:surveys} caption). The numerator
does not depend on the host's intrinsic luminosity: it is set by the beamer
alone and only the noise knows about the star. The factor of two on the
read noise is the difference of two ramp segments. Faint hosts are therefore
{\it better} targets, until the sky and read-noise floor is reached. In GBTDS
fields the noise is dominated by crowding, which correlates with position
rather than host magnitude, and bright-star masking removes some of the
footprint; we adopt a filling factor of $0.7$, so the usable star count is
$1.5\times10^{8}$ rather than the raw $W149<25$ prediction of
$2.1\times10^{8}$ \citep{penny19}. That filling factor is the faint-column search after
bright stars are masked; a giant-host search is a different sample, the
unmasked $W149\simeq14$--$17$ stars, and host shot noise in
Equation~(\ref{eq:snr}) then sets the range (below). $N_{\star}$ in
Equation~(\ref{eq:yield}) is the total $W149<25$ column: the flash does not
need a detected host, and a pointed leak on an unresolved star is still a
detection. The background cut of \S\ref{sec:backgrounds} is stricter. It
asks for coincidence with a resolved, cataloged source. In the bulge those
two populations are almost the same, the mean separation
$0.33^{\prime\prime}$ at $1.22\times10^{8}$~deg$^{-2}$ being comparable to
the \emph{Roman} PSF \citep{penny19}, so the discriminant is weak exactly where the
star-seconds are. At high latitude the cataloged column is the right
$N_{\star}$ for both the yield and the cut, and the discriminant is
strong, but that is not where the harvest is. The faint-host
search is limited by distance, not by host magnitude, and its reach is
\begin{equation}
r_{\rm max}=\left[\frac{\Liso\,\taueff A_{\rm eff}}
{4\pi E_{\gamma}\,Q\,\sigma}\right]^{1/2},
\label{eq:rmax}
\end{equation}
with $\Liso$ the isotropic-equivalent luminosity of the leaked first-ring halo,
$Q$ the adopted threshold and $\sigma$ the matched-filter noise in the
faint-host limit. Relative to a single-sample denominator the matched statistic is
the conservative one for \emph{Roman}'s numbers ($B\simeq50$~e$^{-}$~s$^{-1}$,
$\sigma_{\rm read}\sqrt{n_{\rm pix}}=41.6$~e$^{-}$, $\taueff=60$~s): a bulge
flash of $2.0\times10^{3}$~e$^{-}$~s$^{-1}$ then has matched-filter denominator $80$~e$^{-}$
and ${\rm S/N}=1.5\times10^{3}$. Most GBTDS overlaps are partial
($\taueff\simeq60$~s against a $67$~s exposure; \citealt{penny19,rotac25}), but the bookkeeping is right
on average and ${\rm S/N}\gg Q$ at the bulge, so the yield is safe.

We quote ranges and yields at $Q=8$, which is where the Gaussian trials close.
The number of independent trials is large (\emph{Roman}'s GBTDS
alone supplies $\sim\!1.1\times10^{6}$ reads per star for $\sim\!1.5\times10^{8}$
usable stars, or $1.7\times10^{14}$ read-trials). A matched filter run over a handful of
trial durations multiplies that by a factor of a few, not by $10^{2}$. A
$5\sigma$ threshold would yield $\sim\!5\times10^{7}$ Gaussian false alarms and even
$7\sigma$ leaves a few hundred; at $8\sigma$ the Gaussian expectation is of
order $0.1$ events.

Gaussian trials are not what limits this search, however, and a threshold in
units of $\sigma$ is not the natural statement of its sensitivity. At $Q=8$ the
range is already $r_{\rm max}=109$~kpc on a faint host, past the far edge of the Galaxy, so a
harder cut costs no stars. It costs only the width of the illuminated strip and
the sweep over which the flash stays above threshold: $N_{\rm det}/\fb$ falls
from $0.061$ at $Q=8$ to $0.040$ at $Q=20$ and $0.016$ at $Q=100$, and collapses only near
$Q\simeq1.5\times10^{3}$, where $r_{\rm max}$ drops to the distance of the bulge
(Figure~\ref{fig:limits}c). The Gaussian false-alarm expectation falls by some
seventy decades over that same interval. The threshold should therefore be set
by the measured rate of PSF-coincident instrumental and stellar transients
(\S\ref{sec:backgrounds}) rather than by the Gaussian tail, and the sensitivity
of the search is properly quoted as a pair: the fluence at which the star-offset
control rate is tolerable, and the yield that survives there. The shaded band of
Figure~\ref{fig:limits}c is the decade in $Q$ at which a $10^{-4}$--$10^{-6}$
survivor fraction of the duration-cut sample of \S\ref{sec:backgrounds} would
leave of order one event. For \emph{Roman}
the $8\sigma$ fluence is $640$ detected photons, an average of
$11$~e$^{-}$~s$^{-1}$ over $\taueff$ ($m_{\rm AB}=25.0$ in the F146 photon-rate
zeropoint), against a peak of $2.0\times10^{3}$~e$^{-}$~s$^{-1}$ for a bulge flash. We use
$Q=8$
throughout because that control rate cannot be measured before the data exist,
and Figure~\ref{fig:limits}c gives the yield penalty for any harder cut. The
limits in Table~\ref{tab:surveys} are Poisson limits for a background-free
search; \S\ref{sec:backgrounds} shows those contours are not achievable once
PSF-coincident stellar and instrumental transients are included, and the
estimates there are what would have to be subtracted to reach them.

A $W149=17$ giant at $8$~kpc contributes $\dot N_{\rm src}\simeq1.7\times10^{4}$~e$^{-}$~s$^{-1}$
to the denominator of Equation~(\ref{eq:snr}), so the same flash has
${\rm S/N}=117$ and $r_{\rm max}=31$~kpc. That is still a Galactic range, but it
is not ``visible from anywhere in the Galaxy.'' The dashed curve of
Figure~\ref{fig:limits}c is that giant-host photometry at the full $W149<25$
column and filling factor unity, an upper bound: the true giant subsample is the
magnitude-selected bright tail, whose size we do not assign without a
luminosity function. The $0.7$ fill of Table~\ref{tab:surveys} is the faint-column
search; the giant-host search uses the unmasked bright stars.

\subsection{Expected yields}

The probability $p_{\rm geo}$ that a given transfer illuminates us is the swept solid angle
divided by $4\pi$. Events remain above threshold out to impact parameter
$b_{\rm max}(r)$, the value of $b$ at which $W(b)/W_{\rm max}=(r/r_{\rm max})^{2}$,
so
\begin{equation}
p_{\rm geo}=2\times\frac{\Delta\phi_{\rm eff}\,(2b_{\rm max}\theta_{\rm el})}{4\pi},
\label{eq:pgeo}
\end{equation}
the leading factor accounting for the disjoint launch and arrival leak strips. At
the bulge $b_{\rm max}\simeq4.9$ for \emph{Roman} (the largest $b$ with
$W(b)/W_{\rm max}=(8\,{\rm kpc}/r_{\rm max})^{2}$); at $4$~kpc,
$b_{\rm max}\simeq5.5$ for \emph{Rubin} and $5.4$ for \emph{Euclid} $H_E$.
Because $b_{\rm max}$ is a
function of $r$ it belongs inside the yield integral. We integrate $p_{\rm geo}(r)$
over the disk-plus-bulge column. Evaluating at $8$~kpc instead changes $N_{\rm det}$
by $\sim20$~percent at the fiducial, which we ignore next to the range in piston
residual; near $\vmax\simeq250\kms$, where $r_{\rm max}$ crosses the bulge, the
one-zone value drops discontinuously to zero while the integral still returns a
disk-only harvest of order $0.01\,\fb$. With $R_{\rm L}$ transfers launched per
unit time per system and $T_{\rm obs}$ of open shutter per star, the expected
number of detections is
\begin{equation}
N_{\rm det}=\fb\,N_{\star}(<r_{\rm max})\,R_{\rm L}T_{\rm obs}\,p_{\rm geo},
\label{eq:yield}
\end{equation}
where $\fb$ is the fraction of stars hosting a continuously operating
optical beamer of the kind in \S\ref{sec:beamform}, a $10$~km or $40$~km
array of intercept-matched tiles at a $\lambda/10$ lock, at
$\ms\simeq10^{3}$~kg, $\vmax\gtrsim250\kms$ and $\amax\simeq1$~gee, with
$\lambda$ in $0.3$--$2.3\um$. An ideal lock produces no halo
(\S\ref{sec:beamform}); unlocking the loop saturates at $f_{\rm leak}=0.17$
because the tile Airy still hits the sail. $\amax$ and
$\ms$ are fixed to the tonne-at-a-gee Earth--Mars analog, and $r_{\rm max}$
is a steep function of $\vmax$ (Figure~\ref{fig:design}b). The search is a
null test of that array, which is the practical optical design, not a constraint
on beam-driven light sails as a class. Following
\citetalias{gl15}, where we argued that the infrastructure has high capital and
low operating cost and would be used constantly, it seems reasonable to adopt one
transfer every $6$~h as a reference and to quote limits on
the mean launch rate per star in those units,
$\GamL\equiv 6\,{\rm h}\,\fb R_{\rm L}$. If this is the case, yields scale linearly with
$R_{\rm L}$. At the fiducial rate $N_{\rm det}=0.061\,\fb$, so $N_{\rm det}=1$ wants
$\GamL\simeq16$ and a null search limits $\GamL\lesssim49$: the search
is sensitive in a factor-of-three window of launch rate. The
$\GamL\lesssim49$ contour does not restrict $\fb$ alone, since
$\fb\le1$ by construction. The first-ring level is an upper envelope set by
uniformly illuminated hard-edged tiles, and by the omitted figure and
correlated-piston terms of \S\ref{sec:beamform}; a civilization paying for $1.5$~TW has
reason to apodize, while tile-to-tile amplitude errors and tilt go the other
way.

Table~\ref{tab:surveys} collects the result for a fiducial beamer moving a
$10^{3}$~kg payload at $\vmax=400\kms$ at $1$~gee, a $2.2$~d opposition-class
Earth--Mars crossing rather than a Hohmann analog. Because the intercept-matched pitch scales with the laser wavelength,
$\theta_{\rm el}=1.22\lambda/p$ and therefore $\taueff$ are the same in every
row; the rows differ in collecting area, sampling and stellar column rather than
in the shape of the event. A $1$~m pitch on either envelope is a
tens-of-minutes $m\sim23$ bump and is not this search. Magnitudes are in the
survey photon-rate zeropoint referred to the tabulated $\lambda$ (F146:
$1$~e$^{-}$~s$^{-1}\equiv27.6$~AB); a monochromatic $f_{\nu}$ spread over the
full F146 width is $\simeq0.6$~mag brighter, and we use the photon-rate
convention throughout, so the \emph{Roman}--\emph{Rubin} magnitude offset is not
photon energy alone. Star counts toward the bulge are taken from the Cycle~7 W149$<25$
prediction of \citet{penny19} ($2.40\times10^{8}$ stars in $1.96$~deg$^{2}$, i.e.\
$1.22\times10^{8}$~deg$^{-2}$), rescaled to the $1.7$~deg$^{2}$ GBTDS footprint
recommended by \citet{rotac25}, and assigned a disk-plus-bulge distance
distribution centered at $8.2$~kpc, and multiplied by a filling factor of $0.7$
after bright-star masking. High-latitude columns of $3\times10^{4}$ and
$1.5\times10^{4}$~stars~deg$^{-2}$ are used for the \emph{Rubin} and \emph{Euclid}
footprints \citep{ivezic19,scaramella22}. Figure~\ref{fig:limits} shows the range
as a function of transit speed and the null-search limit as a function of RMS
tile piston.
\afterpage{%
\begin{deluxetable*}{lccccccccc}
\tabletypesize{\scriptsize}
\tablecaption{Sensitivity of the three surveys to light-sail leakage from a
$10$~km or $40$~km array. The halo observables are independent of the envelope
(\S\ref{sec:beamform}); $10$~km is $\sim20\,{\rm kW\,m^{-2}}$, $40$~km the published
directed-energy loading. Beamer
parameters are $\ms=10^{3}$~kg, $\amax=1$~gee, $\vmax=400\kms$, one transfer per
$6$~h, intercept-matched tiles with a $\lambda/10$ lock. Each row is a distinct
beamer: the laser wavelength sets the tile pitch ($p=40$~m at $1.06\um$).
$\lambda=1.06\um$ is the Nd:YAG line, the fiducial we adopt for F146; the other
three rows use the band center. Effective areas are
$A_{\rm eff}=1.95$, $13.6$, $0.60$ and $0.55\,{\rm m^{2}}$. $\taueff$ is the
fluence-equivalent duration at burnout. $m_{\rm pk}$ is the peak flash magnitude
at $8$~kpc. $r_{\rm max}$ is the range of an $8\sigma$ matched-filter detection, and
$\GamL^{95}$ the Poisson upper limit implied by a background-free null search,
for the $\lambda/10$ lock and for a loop-off floor ($f_{\rm leak}=0.17$).
$t_{\rm samp}=22$~s for NISP is one MACC group of the $87$~s photometric ramp.
\S\ref{sec:backgrounds} estimates what would have to be subtracted to reach the
quoted contour; those contours are not achievable once the backgrounds of that
section are included. Euclid yields are $\ll1$ per host fraction, so a null
search does not return a meaningful $\GamL$ limit.
\label{tab:surveys}}
\tablehead{
\colhead{Survey} & \colhead{$\lambda$} & \colhead{$t_{\rm samp}$} & \colhead{$T_{\rm obs}$} &
\colhead{star-s} & \colhead{$\taueff$} & \colhead{$m_{\rm pk}$} & \colhead{$r_{\rm max}$} &
\colhead{$N_{\rm det}/\fb$} & \colhead{$\GamL^{95}$} \\
\colhead{} & \colhead{($\mu$m)} & \colhead{(s)} & \colhead{(s)} & \colhead{} & \colhead{(s)} &
\colhead{(AB)} & \colhead{(kpc)} & \colhead{($\lambda/10$ / off)} & \colhead{($\lambda/10$ / off)}}
\startdata
\emph{Roman} GBTDS (F146)          & 1.06 & 3.04 & $3.4\times10^{6}$ & $5.0\times10^{14}$ & 60 & 19.4 & 109 & $0.061$ / $0.092$ & $49$ / $33$ \\
\emph{Rubin} \emph{LSST} WFD ($r$)        & 0.62 & 30   & $5.5\times10^{3}$ & $3.0\times10^{12}$ & 60 & 18.2 & 145 & $7.4\times10^{-4}$ / $9.0\times10^{-4}$ & $4.1\times10^{3}$ / $3.3\times10^{3}$ \\
\emph{Euclid} EWS ($H_E$)   & 1.50 & 22   & $3.5\times10^{2}$ & $7.3\times10^{10}$ & 60 & 17.5 & 101 & $1.5\times10^{-5}$ / $2.0\times10^{-5}$ & $\cdots$\tablenotemark{a} \\
\emph{Euclid} EWS ($I_E$)   & 0.70 & 566  & $2.3\times10^{3}$ & $4.8\times10^{11}$ & 60 & 19.1 & 71 & $7.5\times10^{-5}$ / $1.1\times10^{-4}$ & $\cdots$\tablenotemark{a} \\
\enddata
\tablenotetext{a}{Yield too low for a meaningful $\GamL$ contour.}
\end{deluxetable*}
}

\subsection{\textit{Roman}}

The Galactic Bulge Time Domain Survey monitors $1.7$~deg$^{2}$ of the bulge in the
wide F146 filter at $\sim\!12$~min cadence with $\sim\!67$~s exposures through six
$\sim\!70$~d seasons \citep{penny19,rotac25}, accumulating $3.4\times10^{6}$~s of
open shutter per star ($39$ days). Sampled at the $3.04$~s read time, this is
$7.0\times10^{14}$ star-seconds at the timescale that matters, two orders of
magnitude more than \emph{Rubin} and four more than \emph{Euclid}, or
$5.0\times10^{14}$ after the filling factor of $0.7$. Combined with
the stellar column toward the bulge ($2.1\times10^{8}$ stars to $W149=25$, of
which $1.5\times10^{8}$ remain after bright-star masking),
\emph{Roman} dominates the monitoring. With the reconstructed $\taueff$ the
$8\sigma$ matched-filter range is $109$~kpc on a faint host; a pointed flash is visible from
anywhere in the Galaxy if the beam points at us. On a $W149\simeq17$ giant,
host shot noise in Equation~(\ref{eq:snr}) brings $r_{\rm max}$ to $31$~kpc. In F146 the photon-rate
zeropoint is $1$~e$^{-}$~s$^{-1}\equiv27.6$~AB, so $m_{\rm pk}=19.4$ at $8$~kpc.
The geometric probability that the beam points at us
is $\Delta\phi_{\rm eff}\times2b_{\rm max}\theta_{\rm el}/2\pi$, i.e.\
$N_{\rm det}\simeq0.061\fb$ at the $\lambda/10$ lock and $N_{\rm det}\simeq0.092\fb$
if the loop is off, so a null search limits $\GamL\lesssim49$
($\lambda/10$) or $\lesssim33$ (loop off), before background subtraction. At the fiducial rate neither contour
cuts $\fb$ below unity.

The GBTDS observes almost entirely in F146, so the color veto of point (ii) is
unavailable at the facility that supplies essentially all of the star-seconds.
This suggests following any \emph{Roman} candidate with
\emph{Euclid}'s simultaneous dichroic (\S\ref{sec:euclid}).

\emph{Roman} is in commissioning and the GBTDS processing configuration is still
open. \emph{Roman} cannot downlink every read and instead averages groups of reads
into resultants on board. The slope recovered from those uneven combinations
\citep{casertano22} degrades only as the square root of the resultant spacing,
so a $\sim\!10$~s effective sampling is entirely workable. What is essential is that the {\it jump information} survive:
the onboard cosmic-ray identification step, which necessarily examines
read-to-read differences, should preserve a compact record of rejected jumps
(position, amplitude, read index) rather than discarding them. At
$\sim5\,{\rm cm^{-2}\,s^{-1}}$ over $\sim300\,{\rm cm^{2}}$ of focal plane and
the $\sim\!2\times10^{7}$~s the WFI integrates over the survey (six pointings
at $3.4\times10^{6}$~s each), that record is a few hundred GB, a negligible
downlink. The computational cost of a PSF-matched fit over
$1.7\times10^{14}$ samples is not negligible; ``no added cost'' here means no
added observing cost. The information is otherwise irrecoverable.

The \emph{Roman} background in Equation~(\ref{eq:snr}) is not a free guess. We
take $7.4$ noise pixels (F146, WFI center) times the \citet{penny19}
season-midpoint sky plus thermal rate
($3.43+1.07$~e$^{-}$~pix$^{-1}$~s$^{-1}$) and add
$\sim\!2$~e$^{-}$~pix$^{-1}$~s$^{-1}$ of residual crowding, giving
$B\simeq50$~e$^{-}$~s$^{-1}$ in the aperture. Zodiacal light toward the bulge is
$2.5$--$7$ times the minimum \citep{penny19}. At $t_{\rm samp}=3.04$~s the
single-read read-noise term ($41.6$~e$^{-}$) dominates $B t_{\rm samp}$ in
variance (a factor $3.4$ in $\sigma$); over $\taueff=60$~s the background and the differenced read
noise are comparable, which is why Equation~(\ref{eq:snr}) is the right
denominator.

\subsection{\textit{Rubin}} \label{sec:rubin}

\emph{Rubin} has the best per-flash sensitivity of the three, by virtue of its
collecting area: at $\vmax=400\kms$ a single halo flash is detected at matched-filter
${\rm S/N}\simeq2.6\times10^{3}$ from $8$~kpc and the range extends to $145$~kpc.
A single $30$~s visit captures at most half of the $60$~s flash, so the S/N that
the cadence actually supplies is $\simeq1.8\times10^{3}$, not the $60$~s matched
filter. Its
limitation is time per star, and the relevant time is single-band. The flash is
monochromatic, so it is caught only while \emph{Rubin} is in the matching
filter: of the $825$ visits of $30$~s the baseline ten-year cadence delivers per
field, the $\sim\!184$ taken in $r$ are the ones that count
\citep{ivezic19}, or $5.5\times10^{3}$~s, $620$ times less than \emph{Roman}.
Even with $5\times10^{8}$
stars in the WFD footprint the yield is $N_{\rm det}\simeq7.4\times10^{-4}\fb$ ($\lambda/10$) or
$9.0\times10^{-4}\fb$ (loop off). Counting all $825$ visits would overstate this by a
factor of $4.5$. \emph{Rubin} is multiband, so the color veto that \emph{Roman}
lacks is available here, although a $30$~s visit is shorter than the $87$~s
doublet and catches one first-ring peak at a time.

Two things would improve this and cost little. First, the current \emph{LSST}
baseline takes single $30$~s exposures; $2\times15$~s snap mode is not in use.
Retaining it in at least part of the survey would halve $t_{\rm samp}$ and
supply a built-in coincidence veto, since a genuine flash appears in one snap
and not the other. Second, single-snap PSF-consistent positive residuals should be
preserved in the alert stream or in a dedicated auxiliary table rather than being
suppressed as artifacts; at present such detections are precisely what
difference-imaging pipelines are tuned to discard.

\subsection{\textit{Euclid}} \label{sec:euclid}

\emph{Euclid}'s reference observing sequence spends only four $566$~s VIS
exposures and four $87$~s NISP exposures per field, so $T_{\rm obs}$ is three to
four orders of magnitude below \emph{Roman}'s. NISP is sampled up the ramp in
MACC$(4,16,4)$; the $22$~s of Table~\ref{tab:surveys} is one of the four groups
that make up the $87$~s integration. With $\taueff\simeq60$~s the NISP
integrations no longer dilute the flash into invisibility ($r_{\rm max}=101$~kpc
in $H_E$), but the yield remains $\ll1$ per host fraction and a null search does
not constrain $\fb$. Because VIS and NISP observe simultaneously through
a dichroic \citep{cropper25,jahnke25}, \emph{Euclid} is the only one of the three
that can establish that a flash present in one band was truly absent in another
{\it at the same instant}, which no sequential multiband survey can do, and which
\emph{Roman} cannot do at all. Any candidate that recurs should be followed up
during an \emph{Euclid} pointing for this reason alone. In addition, if the
beamer's linewidth is intrinsically narrow, a monochromatic flash at
$1.25$--$1.85\um$ would appear in the concurrent NISP red grism as an unresolved
feature on a stellar trace with no continuum counterpart. That search can be run
on existing Q1 and Q2 frames at no observational cost; we do not assume the
linewidth the grism would require.

\section{Backgrounds} \label{sec:backgrounds}

The dominant background is cosmic rays. For \emph{Roman} at L2 the probability
that a cosmic ray lands within a stellar PSF in a given read is $\sim\!10^{-4}$,
so over the GBTDS each star accumulates $\sim\!10^{2}$ coincidences and the
survey as a whole $\sim\!2\times10^{10}$. The discriminant is morphology and duration: a
cosmic ray deposits charge in a contiguous track of one to a few pixels in a
single read, whereas a flash illuminates the full PSF in the correct proportions
for $\sim\!\taueff$. A PSF-matched fit with a $\chi^{2}$ cut across
$\gtrsim10$ pixels, requiring the excess to persist over more than one read,
suppresses ordinary tracks by many orders of magnitude. The residual population is
the extended, quasi-circular events known as snowballs in H4RG and H2RG devices.
On-orbit JWST/NIRSpec at L2 records $0.0015$ snowballs~${\rm cm^{-2}\,s^{-1}}$
for cores of $\gtrsim9$ pixels ($0.3$ per $14.6$~s frame), roughly $100$ times
the ground-test rate \citep{rauscher25}. Scaled to the WFI, that is of order
$10^{7}$ events over the survey, and at a stellar column of
$1.2\times10^{8}$~deg$^{-2}$ essentially every one of them lands in a cataloged
PSF. The deposit itself is a single-read charge dump. A persistence cut that
requires the excess to last more than one read therefore removes the
$10^{7}$ cores; what remains is the fainter post-saturation persistence, not
a $60$~s PSF-shaped flash. The blank-sky control is what measures any
residual. For \emph{Rubin} the
analogous residuals are sub-second satellite and debris glints \citep{corbett20,nir21},
optical ghosts and unmasked bright-star artifacts.

We do not claim these can be eliminated a priori. Instrumental transients are, to
a good approximation, uncorrelated with the positions of cataloged stars, whereas
the signal is by construction coincident with them. In GBTDS fields that
coincidence is with a resolved source (\S\ref{sec:stat}); the yield still
counts the full $W149<25$ column. Running the identical
PSF-matched search at a large ensemble of blank-sky positions therefore yields an
empirical false-alarm rate, with all the non-Gaussianity of the real detector
included and no modeling required. That control fails for the classes most likely
to mimic the signal: persistence from previously saturated sources, inter-pixel
capacitance and cross-talk ghosts at fixed offsets from bright stars,
brighter-fatter residuals, and undersampled-PSF differencing residuals in crowded
fields. In GBTDS fields at $1.2\times10^{8}$ stars~deg$^{-2}$, undersampled
differencing where sources overlap will itself produce PSF-consistent positive
residuals at stellar positions at some rate; the star-offset control partially
handles that blending, and Solar System occultations plus diffraction spikes
are a further, smaller, stellar-position background. A second control, injection-recovery at positions offset by a few
arcseconds from real stars, preserves the local crowding and bright-star
environment while breaking the coincidence. The search is then a measurement of an
excess of PSF-consistent short transients at stellar positions over both the
blank-sky and the star-offset rates. Glints and satellite trails are not perfectly
isotropic, being concentrated near the geostationary belt and in twilight, so the
blank-sky sample should be matched in airmass, hour angle and time of night.

Impulsive spikes in M-dwarf flare continua have been resolved at one-second cadence
\citep{kowalski16}, and a one-second-cadence imaging survey records rise times of
$5$--$100$~s at $\sim\!0.7$~day$^{-1}$ per active M dwarf \citep{aizawa22}. The
GBTDS $W149<25$ column splits as $\sim3.4\times10^{7}$ disk stars and
$\sim1.7\times10^{8}$ bulge stars in $1.7$~deg$^{2}$. Taking $70$ percent of
the disk as M dwarfs and $15$ percent of those as magnetically active gives
$\sim4\times10^{6}$ active dwarfs and
$\sim4\times10^{6}\times0.7\times39\sim10^{8}$ impulsive events over the
open shutter. They are PSF-shaped by construction, so a $\chi^{2}$ cut does
not reject them. A duration cut that requires the excess to be confined to
$\sim\taueff$ and not to persist into the next visit removes classical
hour-scale flares; the Aizawa rise times already sit in the $5$--$100$~s
window, and even a $10$ percent survivor fraction leaves $\sim10^{7}$
events. Quasi-periodic pulsations on M dwarfs occupy the same tens-of-seconds
window: confirmed TESS $20$~s QPPs have periods of a few to ten minutes
\citep{howard22}, and a $17$~percent subset of complex flares is a peak-bump
pair on a minutes-to-hours envelope. Period alone does not discriminate. Duty
cycle does: a QPP is a multi-cycle modulation of a long flare, whereas the leak
is two peaks $87$~s apart and then quiescence, or, for four detections in five, a
single $\sim30$~s peak. Host type does: those flares are disk M dwarfs, not bulge
giants.

We have not shown that an $87$~s pair at a giant position in F146 is
distinct from that class at the $10^{-7}$ level the yield requires. Most of those
hosts are photometrically classifiable as disk dwarfs rather than bulge giants.
A giant-only subsample, $W149\simeq14$--$17$ at $8$~kpc, is a search on unmasked
bright stars with host-noise $r_{\rm max}=31$~kpc; it is not the faint $W149<25$
column of Table~\ref{tab:surveys}, whose $0.7$ fill already masks the same bright
stars. The bulge $W149<25$ column is dwarf-dominated ($W149=25$ is $M\simeq10.5$),
so the giant cut is a magnitude selection, not the disk fraction. What remains is
still large next to $N_{\rm det}\sim0.06\,\fb$, and \emph{Roman} has no color veto.
Sunlit debris and asteroid glints, including the sub-second population
\citep{corbett20,nir21}, land on cataloged stars by chance at a
computable rate. A tens-of-seconds PSF-shaped
brightening of a star is not, by itself, unique; the search is an excess of
single-band, PSF-coincident flashes at giant positions over the disk-dwarf
and star-offset rates. The doublet is the confirmation test for the minority
of chords that clip the axis. Because PSF-coincidence is weak in the bulge
and \emph{Roman} has no color, the discriminants in the fields that carry
essentially all of the star-seconds reduce to duration, morphology and host
type alone.

\section{Discussion and conclusions} \label{sec:disc}

We have presented the optical and near-infrared branch of the light-sail leakage
transient of \citetalias{gl15}. Fresnel matching would shrink the aperture to
$20$--$100$~m at $1\um$ (Equation~\ref{eq:DA}), and that aperture cannot radiate
$1.5$~TW. Spreading that power at $\sim20\,{\rm kW\,m^{-2}}$ is a $10$~km array,
pushing the emitters; Lubin's $10^{3}\,{\rm W\,m^{-2}}$
loading is a $40$~km array, comfortable but aperture-heavy. At fixed $\Ds$, a $\lambda/10$ piston floor on
$40$~m tiles leaks $0.054\,\PA$ into a first-ring halo at
$m_{\rm AB}\simeq19.4$ in F146 at $8$~kpc, with $\taueff\simeq60$~s
(Figure~\ref{fig:beamform}). At intercept match that fraction is
independent of $\PA$ from $10$~GW to $10$~TW, and for a thermally sized
sail so is the peak magnitude (Figure~\ref{fig:power}), so a $10$~GW
plant of the kind committed for AI training is already a Galactic
flash. The typical detection is a single $\sim30$~s peak;
about one in five is a pair $87$~s apart. On a faint host \emph{Roman}'s GBTDS
reaches $r_{\rm max}=109$~kpc at $8\sigma$; on a $W149\simeq17$ giant the range is
$31$~kpc. $N_{\rm det}=1$ wants $\GamL\simeq16$; a null search limits
$\GamL\lesssim49$ ($\lambda/10$) or $\lesssim33$ (loop off), a
factor-of-three window in launch rate, and neither contour restricts
$\fb$ below unity. Neither the reflection nor a later transfer supplies
a second flash. Because $r_{\rm max}$ already exceeds the Galaxy on a faint host,
the threshold is set by PSF-coincident stellar and instrumental transients, and
Figure~\ref{fig:limits}c gives the yield surviving any harder cut.

The limit is a limit on that array and on the tolerance to which it is held,
not on beam-driven light sails as a class. The ramp search does not constrain
metre-class tiling, which leaks a visit-scale $m\sim23$ bump. The recommended
search requires (i) that \emph{Roman} preserve a compact record of rejected ramp
jumps while the GBTDS configuration is still open, (ii) that \emph{Rubin} retain
PSF-consistent single-visit positive residuals, and (iii) that all three be
searched against blank-sky and star-offset controls, with a giant-host cut
against disk-dwarf flares as a separate sample from the faint-column search.
The leaked field of Figure~\ref{fig:beamform}a is an ensemble average. A
single realization is a speckle pattern with grain $\lambda/\DA$, set by
the array envelope, and the pistons are the residual of a live loop, so
that pattern boils on the loop timescale. An observer sweeping a few
$\theta_{\rm el}$ would then see excess variance on top of the $60$~s
envelope. The sail boundary sets the pattern's envelope and the umbra
depth, not the grain. We do not develop that discriminant here: the loop
bandwidth is unmodelled and the contrast depends on it.
Wide-field optical time-domain surveys \citep{ratzloff19,bellm19} and searches
for hour-scale extragalactic flashes \citep{oshikiri24} or century-scale
vanishings \citep{villarroel20} have already covered large sky areas, but none
has the bulge stellar column, the star-coincidence requirement, and the
tens-of-seconds ramp sampling of the GBTDS together. If intensity-limited
optical beamers of this kind are commonly employed in our galaxy, this activity
could be revealed by a PSF-matched ramp search of \emph{Roman}, \emph{Rubin} and
\emph{Euclid} at no additional observing cost. Nearby multiply-transiting
systems near projected conjunction, the strategy we proposed in
\citetalias{gl15} and the window \citet{jwright18} takes for intercepting
leaked propulsion once the orbits are known, remain the right targets for
dedicated instruments; the
surveys' statistical reach is over $10^{8}$--$10^{9}$ distant stars.

\begin{acknowledgments}
This work was supported in part by the \emph{Galileo Project}.
We thank the \emph{Roman}, \emph{Rubin} and \emph{Euclid} project teams for making
detailed survey specifications publicly available. We thank J. Xavier Prochaska for providing
access to AI models that were used to conduct independent AI reviews.
This manuscript was prepared
with the assistance of the Anthropic AI model Claude Opus~5 and X.ai model Grok 4.6. All text
that was not written by the authors was reviewed and approved by the authors, and
all references have been verified.\footnote{The \LaTeX\ source and figures are available at
\url{https://github.com/guillochon/light-sails-followup}.}
\end{acknowledgments}

\software{astropy \citep{astropy22}, numpy \citep{numpy20}, scipy \citep{scipy20},
matplotlib \citep{matplotlib07}}

\bibliographystyle{aasjournalv7}
\bibliography{bibliography}

\end{document}